\documentclass[conference]{IEEEtran}
\IEEEoverridecommandlockouts
\usepackage{cite}
\usepackage{amsmath,amssymb,amsfonts}
\usepackage{algorithmic}
\usepackage{graphicx}
\usepackage{textcomp}
\usepackage[dvipsnames]{xcolor}
\usepackage[inline]{enumitem}
\usepackage[a4paper, total={184mm,239mm}]{geometry}
\usepackage{siunitx}
\usepackage{soul}

\usepackage{tikz}
\usetikzlibrary{shapes,arrows,positioning,calc}
\usetikzlibrary{arrows.meta, quotes, patterns, angles}

\usepackage{titlesec}
\titlespacing*{\section}
{0pt}{1.2ex plus 1ex minus .2ex}{0.8ex plus .2ex}
\titlespacing*{\subsection}
{0pt}{1.2ex plus 1ex minus .2ex}{0.8ex plus .2ex}
\titlespacing*{\subsubsection}
{0pt}{1.2ex plus 1ex minus .2ex}{0.8ex plus .2ex}

\usepackage{hyperref}

\def\BibTeX{{\rm B\kern-.05em{\sc i\kern-.025em b}\kern-.08em
    T\kern-.1667em\lower.7ex\hbox{E}\kern-.125emX}}

\newcommand{\jun}[1]{{\textcolor{black}{#1}}}

\newcommand{\review}[2]{\textcolor{black}{#2}}
\newcommand{\sdamiano}[1]{\textcolor{black}{#1}}
\makeatletter
\newcommand{\newlineauthors}{%
  \end{@IEEEauthorhalign}\hfill\mbox{}\par
  \mbox{}\hfill\begin{@IEEEauthorhalign}
}
\makeatother

\IEEEaftertitletext{\vspace{-1.4\baselineskip}}

\begin{document}

% Jun: set the space between different bodies
\setlength{\intextsep}{0.cm}
\setlength{\textfloatsep}{1.0ex}
\setlength{\floatsep}{1.0ex}
\setlength{\abovecaptionskip}{0.1cm}
\setlength{\belowcaptionskip}{0.1cm}

\title{Real-Time Acoustic Perception for Automotive Applications}

% \author{\IEEEauthorblockN{1\textsuperscript{st} Jun Yin}
% \IEEEauthorblockA{\textit{Dept. of Electrical Engineering ESAT-MICAS} \\
% \textit{KU Leuven}\\
% Leuven, Belgium \\
% jun.yin@esat.kuleuven.be}
% \and
% \IEEEauthorblockN{2\textsuperscript{nd} Stefano Damiano}
% \IEEEauthorblockA{\textit{Dept. of Electrical Engineering ESAT-STADIUS} \\
% \textit{KU Leuven}\\
% Leuven, Belgium \\
% stefano.damiano@esat.kuleuven.be}
% \and
% \IEEEauthorblockN{3\textsuperscript{rd} Andre Guntoro }
% \IEEEauthorblockA{\textit{Robert Bosch GmbH} \\
% Renningen, Germany \\
% andre.guntoro@de.bosch.com}
% \newlineauthors
% \IEEEauthorblockN{4\textsuperscript{th} Marian Verhelst}
% \IEEEauthorblockA{\textit{Dept. of Electrical Engineering ESAT-MICAS} \\
% \textit{KU Leuven}\\
% Leuven, Belgium \\
% marian.verhelst@kuleuven.be}
% \and
% \IEEEauthorblockN{5\textsuperscript{th} Toon van Waterschoot}
% \IEEEauthorblockA{\textit{Dept. of Electrical Engineering ESAT-STADIUS} \\
% \textit{KU Leuven}\\
% Leuven, Belgium \\
% toon.vanwaterschoot@esat.kuleuven.be}
% }

\author{
\IEEEauthorblockN{Jun Yin\IEEEauthorrefmark{1}\textsuperscript{\textsection}, Stefano Damiano\IEEEauthorrefmark{2}\textsuperscript{\textsection}, Marian Verhelst\IEEEauthorrefmark{1}, Toon van Waterschoot\IEEEauthorrefmark{2}, Andre Guntoro\IEEEauthorrefmark{3}$^\star$}
\IEEEauthorblockA{\textit{\IEEEauthorrefmark{1}ESAT-MICAS KU Leuven, \IEEEauthorrefmark{2}ESAT-STADIUS KU Leuven, \IEEEauthorrefmark{3}Robert Bosch GmbH}}
% \IEEEauthorblockA{\review{}{Email: \{jun.yin, stefano.damiano\}@esat.kuleuven.be, ...}}%\textcolor{orange}{Mail address? Co-first Authorship in DATE? Corresponding author to whom?}
\IEEEauthorblockA{$^\star$Email: andre.guntoro@de.bosch.com}
\thanks{\textsection\;These authors contribute equally to the paper.}
}

\maketitle

\begin{abstract}
%\mv{I beleve all DATE submissions are blind, and we hence have to remove the author list. Also merge all affilitions in 1 line, to take less space. E.g. see here: \href{url}{https://ieeexplore.ieee.org/stamp/stamp.jsp?arnumber=9474107&casa_token=0Uo2PEgcFrIAAAAA:Orm02y9EGbPSCJEkK2hFAa3Hc4oWOVTPbgqYCZnZyQhpiABOSj3iDVef6YKV2gNw65c67s8AOj8}\\}
In recent years the automotive industry has been strongly promoting the development of smart cars, equipped with multi-modal sensors %that enable them 
to gather information about the surroundings, in order to aid human drivers or make autonomous decisions. While the focus has mostly been on visual sensors, also acoustic events are crucial to detect situations that require a change in the driving behavior, such as a car honking, or the sirens of approaching emergency vehicles. 
In this paper, we summarize the results achieved so far
% and the challenges ahead 
in the Marie Sk\l{}odowska-Curie Actions (MSCA) Eruopean Industrial Doctorates (EID) project "Intelligent Ultra Low-Power Signal Processing for Automotive (I-SPOT)".
On the algorithmic side, the I-SPOT Project aims to enable detecting, localizing and tracking environmental audio signals by jointly developing microphone array processing and deep learning techniques that specifically target automotive applications. Data generation software has been developed to cover the I-SPOT target scenarios and research challenges. This tool is currently being used to develop low-complexity deep learning techniques for emergency sound detection. On the hardware side, the goal impels workflows \review{}{for hardware-algorithm} co-design to \review{}{ease the generation of architectures that are sufficiently flexible towards algorithmic evolutions without giving up on efficiency, as well as enable rapid feedback of hardware implications of algorithmic decision.} %into the algorithm tuning stage, as well as the rapid hardware modeling and profiling for algorithm updates. 
This is pursued though a hierarchical workflow that breaks the hardware-algorithm design space into reasonable subsets, which has been tested for operator-level optimizations on state-of-the-art robust sound source localization for edge devices.
% \mv{Ending sentence on targets?}
Further, several open challenges towards an end-to-end system are clarified for the next stage of I-SPOT.
\end{abstract}


\section{The I-SPOT Project}
\label{sec:introduction}
\begin{figure}[t]
    \centering
    \resizebox{0.85\linewidth}{!}{
        \includegraphics{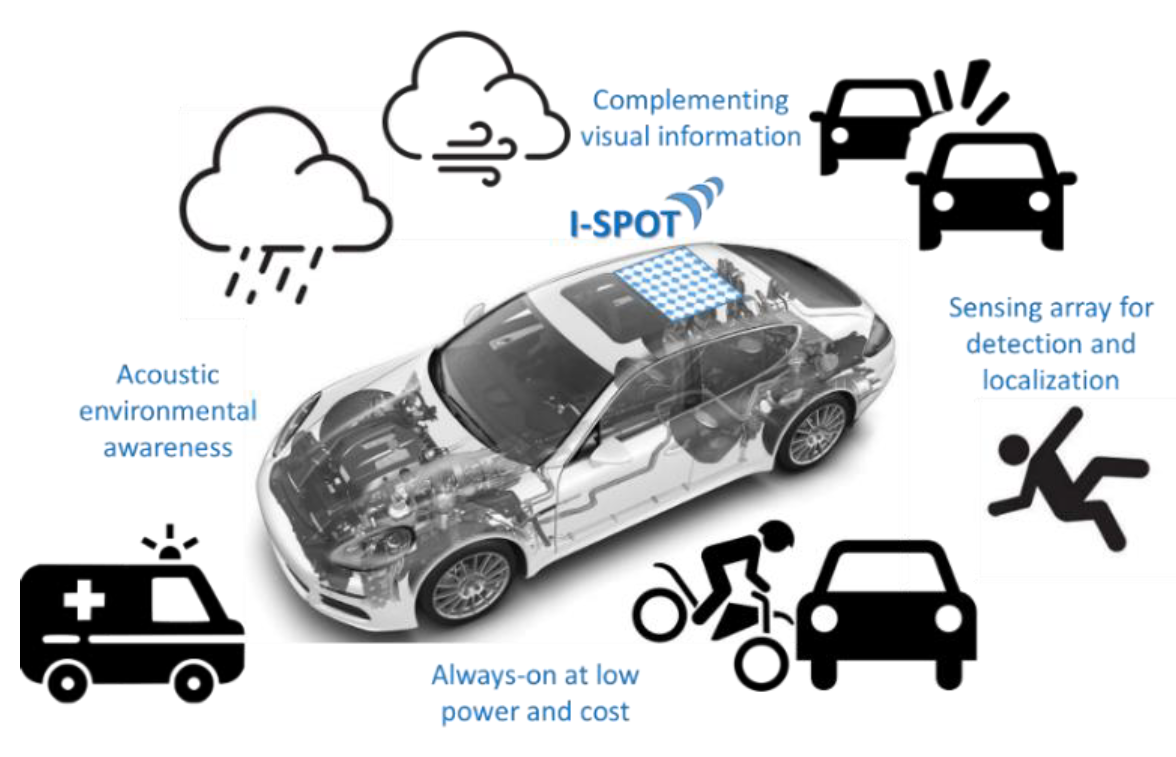}
    }
    \caption{I-SPOT targets the enhancement of smart cars with acoustic environmental awareness, through the addition of a smart sensing array capable of acoustic localization and detection, supported by custom processing hardware for efficient always-on operation.}
    \label{fig:i-spot_target}
\end{figure}

The automotive industry is currently going through radical innovations that require rethinking the concept of mobility and the role of new technologies in the development of the automobile of the future. A key role in this evolving context is played by smart vehicles, aiming on one side to provide assistance to human drivers by enhancing their environmental awareness or supplying useful information in real-time, and on the other hand to achieve some level of autonomy in taking decisions and (partially) controlling the driving.
At the time of writing, the first vehicles reaching level 3 of driving automation\cite{iso/sae_pas_22736_taxonomy_2021} begin to appear on the market, while manufacturers strive to achieve higher autonomy in future years, fueling the development of reliable technical solutions to ensure the safety of automated driving systems. Autonomous cars extensively rely on perceptual abilities, i.e. on the ability to extract in real time information that is useful for driving safely, from data collected via different and multi-modal sensors mounted on the car.
Most of the research on how to enhance the situational awareness of vehicles has been focusing on visual scene analysis, exploiting computer vision techniques and both long- and short-range imaging devices such as Light Detection and Ranging sensors (LiDARs), radars and cameras\cite{hussain_autonomous_2019}. Human drivers, however, also strongly rely on \emph{acoustic} cues as a way to perceive and understand what is happening in their surroundings: in certain scenarios, acoustic events are not accompanied by corresponding visual cues (e.g. a car honking) while in others they can enable the detection of objects that are invisible due to occlusions (e.g. an emergency vehicle behind a corner with an active siren). Acoustic perception can therefore complement the knowledge obtained from other sensory devices both in active (\emph{drive}) and passive (\emph{park}) mode, where sensing abilities can be exploited to continuously monitor the car and the surrounding environment for possible hazards (\emph{always-on} operation). Among multiple potential use cases\cite{marchegiani_how_2022}, highlighted in Fig.~\ref{fig:i-spot_target}, such technology could help in:
\begin{enumerate*}[label=(\roman*)]
    \item detecting dangerous situations;
    \item identifying anomalies in car components;
    \item monitoring the acoustic scene for critical events.
\end{enumerate*}

To compensate for the current lack of acoustic awareness in cars and enhance the perception ability of autonomous vehicles, the EU MSCA project I-SPOT \cite{i-spot_intelligent_2021} targets the exploitation of acoustic sensing in the automotive domain. The project started in Nov. 2020 and relies on the collaboration between KU Leuven  and Robert Bosch GmbH.
Two early-stage researchers are involved in conducting research at the two partner institutions, working respectively on the algorithmic challenges and on the hardware design, with the ultimate goal of merging the two components within this project. The project is now in its intermediate stage, and will run until Oct. 2024.

In this paper we describe the I-SPOT Project by introducing its motivation and goals, the results achieved during the first project stage and the problems still to be addressed. The rest of the paper is organized as follows: in Sec.~\ref{sec:project_goals} we introduce the target goals of I-SPOT, both on the algorithmic side and on hardware design aspects.  Sec.~\ref{sec:related_work} discusses the related research constituting the state of the art at the start of the project. Next, in Sec.~\ref{sec:achievements} we present what has been achieved so far and in Sec.~\ref{sec:open_challenges} the work that still needs to be done during the second stage of the project. We finally draw conclusions in Sec.~\ref{sec:conclusion}.

\section{Project Goals}
\label{sec:project_goals}
The ultimate goal of the I-SPOT Project is the enhancement of perceptual abilities of smart cars via acoustic sensing and signal processing. This broad task incorporates many sub-problems that can be grouped into algorithmic development and hardware design, tackled by the two researchers in an interleaved manner, where hardware assessment becomes an influential part of algorithm design and vice versa.

% \textcolor{blue}{[JUN] Move the "On the algo level" to start from the next paragraph. Use this paragraph to elaborate automotive challenges (Reviewer-2)?}
On an algorithmic design level, the project targets the development of low-footprint signal processing techniques for automotive acoustic signal identification, characterization and localization. This task corresponds to an application of the well-known Sound Event Localization, Detection (and Tracking), or SELD(t) problem in the automotive domain\cite{adavanne_localization_2019}. This use case brings several application-specific challenges to be addressed in the project: 
% \textcolor{blue}{[JUN] Look like there are chances to shrink these bullets if we use this discussion as general automotive challenges and save space? (E.g. point-1 intertwined source\&noise classes and challenging SNR [abbr first usage if you have it here]; point-2 safety\&accuracy; point-3 robust dataset\&training; [Then 1/2/3 can be even enumerated in one bullet with example cases ...]; point-4/5 I can shrink more as it is a bit overlapping with my bullets afterwards;)}
\begin{enumerate}
    \item The automotive acoustic scene is characterized by strong and dynamic background noise generated by multiple overlapping sound sources (e.g. other vehicles, people, environmental sounds such as rain or wind, etc.);
    \sdamiano{\item Due to the foreseen safety-critical use case, a high accuracy is required in both the detection and localization of highly variable sound events (e.g. siren sounds are usually different in each country or region). Thus, algorithms should be robust to noise and interference while having a strong generalization ability;}
    % \item Due to the foreseen safety-critical use case, a high accuracy is required in both the sound event detection and localization;
    % \item The algorithms should be robust to the high variability of the sound events targeted in automotive applications (e.g. siren sounds are usually different in each country or region);
    \item The target deployment on embedded devices requires the algorithms to be multi-mode and computationally efficient, i.e., including the fully-functional low-latency driving mode and trigger-based low-power parking mode;
    % \item The target deployment on embedded devices requires the algorithms to be computationally and power-efficient: when operating in drive mode, a low-delay output is demanded, while an ultra-low power consumption is essential for the use in park mode;
    \item To fulfill such efficiency, algorithm development and hardware design should form iterations, namely the software-hardware co-design flow, where rapid design reuse and hardware-algorithm co-optimization can take place.
    % \item To enable such efficient execution, algorithms cannot be developed in isolation from the embedded hardware design. Specifically, a hardware-algorithm co-design flow is required, which enhances the collaboration between the project partners.
\end{enumerate}

Although the SELD(t) problem has been discussed at length in the literature and solutions have been proposed for both indoor (e.g. speaker identification \cite{he_deep_2018}, anomalous sound detection\cite{mnasri_anomalous_2022}) and outdoor applications (e.g. acoustic scene classification\cite{barchiesi_acoustic_2015}, traffic monitoring\cite{mnasri_anomalous_2022}), existing solutions do not cover these specific challenges of the automotive domain. Accommodating them constitutes one of the core innovative aspects of the I-SPOT Project. 

%The second focus of the algorithmic-related research \mv{[Instead of defining this as an algorithmic challenge, should we not define the sensor array design as a system challenge? E.g. look at fig 2 of our proposal. We can maybe even include this figure?]} is the study of an optimal microphone array topology and {its optimal} placement on the car body, in order to improve the quality of the received signal and boost the performance of the detection and localization algorithms. This task requires the definition of the desired number of sensors and their relative position, two parameters that can strongly influence the localization algorithms~\cite{dmochowski_generalized_2007}. While the problem has been addressed in room scenarios, few works concerning the automotive domain~\cite{barak_microphone_2020} exist. In such applications, only few places can effectively be used, due to the requirements of car manufacturers and the necessity to protect the sensors from the harsh environment, where the presence of strong vibrations, wind, temperature changes and atmospheric phenomena could cause damage and harm their correct operation.

% \jun{Background in hw deployment considerations}
% \jun{deployment on edge devices -> current challenges}
\jun{
Towards the hardware implementation and deployment, these algorithm targets also pose significant challenges. 
Firstly, hardware efficiency is critical in executing aforementioned computationally intensive algorithms in resource-restricted edge devices. Currently, the thriving field of AI deployment provides a wide range of hardware solutions for neural network execution \cite{reuther2021ai}. However, the targeted algorithms require a hybrid approach with both neural networks in combination with conventional signal pre-/post-processing. Hardware platforms efficiently supporting such heterogeneous, diverse workloads are scarce and needed for power-efficient, low-latency end-to-end acoustic solutions \cite{li2022recent}. Moreover, the automotive requirements ask for a reliable and well-packaged hardware system.
Hence, on the hardware level, the I-SPOT system needs to feature:
\begin{enumerate}
    \item Multi-kernel support for end-to-end 
    % heterogeneous
    hybrid
    algorithms;
    \item Real-time low-latency operation to quickly response to % since the commencement of 
    each target event;
    \item An optimized energy efficiency, to reduce the overall automobile power budget, especially in park mode.
\end{enumerate}
}

\jun{
Besides striving for efficiency under heterogeneous workloads, a second important goal of I-SPOT's hardware development is to enable rapid design iterations and tolerate later algorithmic changes. In contrast to a traditional accelerator design flow in which the algorithm is considered stable when the accelerator development starts, the I-SPOT algorithm will evolve strongly during and even after hardware generation. %On the one hand, although communities like the DCASE have been trailblazing for better SELD solutions, low-complexity optimization and design parameter sweet-spot remains open for the I-SPOT target scenario. 
We hence have to avoid on one hand that dedicated hardware architecture would reject new algorithmic features during the project, yet, also shy away from the development of too generic, flexible hardware architectures which lead to a lack of efficiency. This hence introduces several requirements for the I-SPOT hardware design workflow:
\begin{enumerate}
    \item It is necessary to find a balance between flexibility and application-specific hardware architectures, to provide agile support of future algorithm upgrades;
    \item \review{}{The design procedure should enable rapid feedback for 
    % algorithm optimization via hardware assessment
    hardware-algorithm co-optimization
    among multiple hardware design levels, especially the early design stage;} % Jun: mv: cost modeling and assessment of algorithmic impact (more algorithm HW/alg co-design related.
    \item \review{}{The workflow needs to feature hardware design reuse for fast integration and incremental module optimization;}
    \item \review{}{Software-related optimization passes ought to form a scheduling tool that offers user-friendly programming of the resulting hardware.}
    % The workflow needs to realize design reuse and rapid evaluations that support co-design features such as the hardware-algorithm design space exploration.  
    % \mv{[I do not understand the difference between nb 2 and 3? I would make a separate bullet on design reuse (is hardware related, maybe combine with speed of design, automated synthesis, etc.) and on cost modeling and assessment of algorithmic impact (more algorithm HW/alg co-design related.  Do we not also need a bullet on ease of programming/scheduling/compilation?]}
\end{enumerate}
}

The last system-level challenge is related to the assessment of the optimal microphone array topology and placement on the car body, in order to improve the quality of the received signal and boost the performance of the detection and localization algorithms. This task requires the definition of the desired number of sensors and their relative position, two parameters that can strongly influence the localization algorithms~\cite{dmochowski_generalized_2007}. While the problem has been addressed in room scenarios, few works concerning the automotive domain~\cite{barak_microphone_2020} exist. In such applications, only few places can effectively be used, due to the requirements of car manufacturers and the necessity to protect the sensors from the harsh environment, where the presence of strong vibrations, wind, temperature changes and atmospheric phenomena could cause damage and harm their correct operation.

To summarize, the I-SPOT Project embraces the design of an end-to-end system to provide autonomous cars with acoustic perception abilities.
% The foreseen challenges range from the definition of the microphone array architecture, to the joint development of algorithms for the detection and localization of emergency sounds, and design of custom edge-devices for their deployment and real-time execution. 
The foreseen challenges elaborated in this section link to a wide range of related research addressing the microphone array topology, the detection and localization of critical events, and efficient agile hardware architecture.

\section{Related Work}
\label{sec:related_work}
Despite the wide literature on acoustic scene analysis, works targeting automotive applications are still limited. Some authors have tackled the detection and localization of emergency sound events, such as car horns and emergency sirens in an urban scenario~\cite{tran_acoustic-based_2020, cantarini_acoustic_2021, marchegiani_listening_2022, walden_improving_2022, sharma_emergency_2021, furletov_auditory_2021, sun_emergency_2021}. These works can be split into the ones addressing only the detection problem~\cite{tran_acoustic-based_2020, cantarini_acoustic_2021,walden_improving_2022, sharma_emergency_2021} and the ones targeting localization as well~\cite{marchegiani_listening_2022, sun_emergency_2021, furletov_auditory_2021}. 
The proposed solutions to both problems are mostly based on end-to-end machine learning and deep learning methods, that have proven to provide an increased robustness to strong background noise and complex dynamic acoustic scenes as compared to traditional signal processing techniques~\cite{marchegiani_listening_2022}. The state-of-the-art approaches share a similar processing pipeline for the detection stage, eventually followed by a localization stage. The detection process can be summarized as follows. First, a feature-extraction step takes as input the audio signal recorded by one (for the detection-only problem) or multiple (if the localization is addressed too) microphones and builds a corresponding representation to be used as input to a neural network.
Most methods use time-frequency feature representations, such as spectrograms~\cite{cantarini_acoustic_2021, tran_acoustic-based_2020,walden_improving_2022, sun_emergency_2021}, gammatonegrams~\cite{cantarini_acoustic_2021, marchegiani_listening_2022}, Mel-frequency cepstral coefficients (MFCCs)~\cite{tran_acoustic-based_2020,sharma_emergency_2021,sun_emergency_2021}, or, less commonly, gammatone-frequency cesptral coefficients (GFCCs), constant-Q transform (CQT) and chromagrams~\cite{sharma_emergency_2021}. Others take the raw waveform of the windowed audio signal as input feature~\cite{furletov_auditory_2021}. Finally, some approaches adopt both time-frequency representations and the raw-waveform feature to train multi-path neural networks ~\cite{tran_acoustic-based_2020, sun_emergency_2021}. 
After the feature extraction phase, the techniques for sound event detection and classification are mostly based on Convolutional Neural Networks (CNNs)~\cite{cantarini_acoustic_2021, tran_acoustic-based_2020,walden_improving_2022,sun_emergency_2021, sharma_emergency_2021}, with the exception of~\cite{furletov_auditory_2021}, exploiting a fully-connected neural network, and~\cite{marchegiani_listening_2022}, based on a U-net architecture~\cite{ronneberger_u-net:_2015} for the detection stage. The localization, instead, is tackled in~\cite{sun_emergency_2021} jointly with the detection, using an additional direction of arrival output added to the same network, while~\cite{marchegiani_listening_2022} includes a second CNN that takes as input the segmented gammatonegram features obtained during the detection stage. In~\cite{furletov_auditory_2021}, a traditional signal processing stage is cascaded to the detection network, in order to estimate both the sound's direction of arrival and distance.

% SDamiano: TRACK THE NETWORK QUANTIZATION REFERENCES AND SEE IF THEY CAN BE IMPROVED
% In the existing literature, the main focus is posed
% while considerations on the model complexity, the detection delay, and the optimization of the model for the deployment on edge-devices for real-time applications are rarely discussed.
%\jun{Although edge-level solutions have been extensively studied in similar fields such as the automatic speech recognition \cite{malik2021automatic}, the I-SPOT target domain mainly focuses on the accuracy of the detection and localization.} \mv{[Previous sentence seems to be out of place, given the previous paragraph, which already discusses detection and localization?]} 
Towards the realization of efficient models for the edge, the Detection and Classification of Acoustic Scenes and Events (DCASE) community \cite{dcase_challenge_2022} has recently obtained sub-100K-parameter models for low-complexity acoustic scene classification (DCASE2022 task-1), while the SELD problem (DCASE2022 task-3) still relies on over 10M neural network weights. 
The I-SPOT Project embodies such considerations in the algorithmic design, via the introduction of a co-design workflow to promote the joint optimization of model accuracy and complexity. This new research trend is still weakly explored in the audio signal processing research community~\cite{mohaimenuzzaman_pruning_2022, cerutti_sound_2020} as compared to the well-established research on network quantization and pruning for image processing using deep learning techniques~\cite{gholami_survey_2021}. This requires strengthening the interconnection between the heterogeneous algorithm development exploiting deep learning in combination with traditional signal processing on one hand, and versatile reconfigurable hardware architectures optimized for the automotive setting on the other hand. %embedded AI design processes and the deep learning techniques for signal processing, leading to innovative algorithmic design procedures 

\jun{% Relavent Hardware in CGRA and multi-core\\design iteration from overlay and IR choices
%The co-design workflow requires the hardware prototype to be modular and tunable for design iterations. 
The most straightforward approach towards hardware versatility across a wide range of algorithms is to make the hardware fully programmable. General-purpose devices such as CPUs and GPUs satisfy the demand of computational performance and runtime programmability for end-to-end hybrid algorithms, yet lack energy efficiency. More specialized devices like field-programmable gate arrays (FPGAs) could be controlled and configured at finer-granularity, yet still suffer from low power efficiency, and require considerable programming time. To overcome these bottlenecks, the modern coarse-grained re-configurable arrays (CGRA) \cite{podobas2020survey} promise to offer a better balance between processor flexibility and energy efficiency. Several open-sourced CGRA platforms \cite{karunaratne2017hycube, bahr2020creating, anderson2021cgra, tan2021opencgra} have been proposed to accelerate heterogeneous workloads within certain application domains with more flexibility through run-time conditional controls. CGRA-based scheduling is explored at both fine and coarse hardware granularity, targeting to achieve either %compatible for the combination with other micro-architecture to achieve 
optimal utilization \cite{karunaratne2017hycube} through low level reconfigurability, or ultra-low power operation \cite{gobieski2021snafu, giordano2021chimera} through specialized heterogeneous fabric design. Yet, the mapping algorithms for CGRAs remain challenging, and still fail to smoothly compile applications to modern CGRAs, especially when complex CGRA fabrics scale up.
}

\section{Achievements}
\label{sec:achievements}
% SDamiano: CHECK IF DATA GENERATION IS MENTIONED AS A PROJECT GOAL
During the first stage of the project, preliminary objectives have already been met, laying the foundations for the research tasks in the second phase towards the target goals presented in Sec.~\ref{sec:project_goals}. 
\subsection{Algorithmic Achievements}
\label{subsec:algorithmic_achievements}

The first problem that has been addressed is the generation of data to design and assess algorithms for sound source detection and localization. Although some datasets containing urban sounds are available in the literature~\cite{salamon_dataset_2014, piczak_esc:_2015, gemmeke_audio_2017}, they can hardly be used to address the I-SPOT challenges. First, these datasets target only the \emph{detection} problem, thus usually providing single channel recordings with temporal labels. To perform sound source localization, instead, spatial information about the position and movement of sound sources in the acoustic scene is needed, and multi-microphone array data are required. 
Second, a tool enabling to flexibly generate data while changing the array configuration is of utmost importance to analyze the impact of different microphone array architectures on the detection and localization performance. This research topic is often missing in the literature, particularly in the automotive use case, probably also owing to the unavailability of sufficient data to support a systematic assessment.

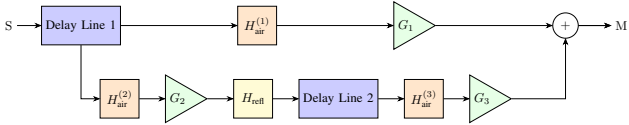
\begin{figure}[t]
    \centering
    \resizebox*{0.95\linewidth}{!}{
        \tikzset{
    block/.style = {draw, rectangle, minimum height = 3em, minimum width = 3em},
    sum/.style= {draw, fill=white, circle},
    gain/.style = {draw, isosceles triangle, minimum height = 1em,     isosceles triangle apex angle=60},
    port/.style = {inner sep=0pt, font=\tiny},
    input/.style = {coordinate}, % Input
    joint/.style = {circle, draw, fill, inner sep=0pt, minimum size=2pt},
}

% \begin{tikzpicture}
%     \node [name = rinput] (rinput) at (0.0,0) {SRC};
%     \node [input,name=input1,thick,above] at (0,0){};
%     \node[block] (delayline1) at (2,0) {Delay Line 1};
%     \node[block] (airabs1) at (4,0) {$H_{air}^{(1)}$};
%     \node[gain] (gain1) at (6,0) {$G_1$};
%     \node[sum] (sum) at (14,0) {$\Sigma$};
    
%     \node[block] (airabs2) at (3,-2) {$H_{air}^{(2)}$};
%     \node[gain] (gain2) at (5,-2) {$G_2$};
%     \node[block] (asphalt) at (7,-2) {$H_{refl}$};
%     \node[block] (delayline2) at (9,-2) {Delay Line 2};
    
%     \node[block] (airabs3) at (10,-4) {$H_{air}^{(3)}$};
%     \node[gain] (gain3) at (12,-4) {$G_3$};
    
%     \node (out) at (15,0) {};
%     \node at (15.5,0)  {MIC};
    
%     \draw[->] (rinput) -- node {}(delayline1);
%     \draw[->] (delayline1) -- node {} (airabs1);
%     \draw[->] (airabs1) -- node{} (gain1);
%     \draw[->] (gain1) -- node {} (sum);
%     \draw[->] (sum) -- node {} (out);
    
%     \draw[->] (delayline1) |- node {} (airabs2);
%     \draw[->] (airabs2) -- node{} (gain2);
%     \draw[->] (gain2) -- node{} (asphalt);
%     \draw[->] (asphalt) -- node{} (delayline2);
    
%     \draw[->] (delayline2) |- node {} (airabs3);
%     \draw[->] (airabs3) -- node {} (gain3);
%     \draw[->] (gain3) -| node {} (sum);
%     \draw[->] (sum) -- node {} (out);
    
% \end{tikzpicture}

\begin{tikzpicture}[>=Stealth]
        \node [name = rinput] (rinput) at (0.0,0) {$\mathrm{S}$};
        \node [input,name=input1,thick,above] at (0,0){};
        \node[block, fill = blue!20, right = 2.5em of input1] (delayline1) {Delay Line 1};
        \node[block, fill = orange!20, right = 9em of delayline1] (airabs1) {$H_\text{air}^{(1)}$};
        \node[gain, fill = green!10, right = 9em of airabs1] (gain1) {$G_1$};
        \node[sum, right = 9em of gain1] (sum){$+$};
        
        \node[block, fill = orange!20] (airabs2) at (3,-2) {$H_\text{air}^{(2)}$};
        \node[gain, fill = green!10, right = 2em of airabs2] (gain2) {$G_2$};
        \node[block, fill = yellow!20, right = 2em of gain2] (asphalt) {$H_\text{refl}$};
        \node[block, fill = blue!20, right = 2em of asphalt] (delayline2) {Delay Line 2};
        
        \node[block, fill = orange!20, right = 2em of delayline2] (airabs3) {$H_\text{air}^{(3)}$};
        \node[gain, fill = green!10, right = 2em of airabs3] (gain3) {$G_3$};
        
        \node[right = 2.5em of sum] (out) {};
        \node[right = -0.7em of out] {$\mathrm{M}$};
        
        \draw[->] (rinput) -- node {}(delayline1);
        \draw[->] (delayline1) -- node {} (airabs1);
        \draw[->] (airabs1) -- node{} (gain1);
        \draw[->] (gain1) -- node {} (sum);
        \draw[->] (sum) -- node {} (out);
        
        \draw[->] (delayline1) |- node {} (airabs2);
        \draw[->] (airabs2) -- node{} (gain2);
        \draw[->] (gain2) -- node{} (asphalt);
        \draw[->] (asphalt) -- node{} (delayline2);
        
        \draw[->] (delayline2) -- node {} (airabs3);
        \draw[->] (airabs3) -- node {} (gain3);
        \draw[->] (gain3) -| node {} (sum);
        \draw[->] (sum) -- node {} (out);
        
    \end{tikzpicture}
    }
    \caption{\emph{Pyroadacoustics} block scheme. Acoustic propagation is modeled using variable-length delay line elements; the sound attenuation caused by spherical propagation is implemented via three gain elements $G_1, G_2, G_3$; the asphalt reflection properties are modeled using an FIR filter $H_\mathrm{refl}$; the air absorption properties are modeled via three FIR filters $H_\mathrm{air}$~\cite{damiano_pyroadacoustics:_2022}.}
    \label{fig:pyroad_architecture}
    
\end{figure}

\begin{figure}[t]
    \centering
    \resizebox*{0.6\linewidth}{!}{
        \begin{tikzpicture}[>=Stealth]
    
    % Road Surface
    \draw (0,0) --++(7,0);
    \fill[pattern=north east lines] (0,-0.3) rectangle ++(7,0.3); 

    % Source and Receiver
    % \draw (1,2) node[anchor = south, label = S, name = s] {\textbullet};
    % \draw (5,1.5) node[anchor = south, label = R, name = r] {\textbullet};

    \node[circle, fill = black, inner sep=0pt,minimum size=4pt,label= $\mathrm{S}$] (s) at (1,2) {};
    \node[circle, fill = black, inner sep=0pt,minimum size=4pt,label= $\mathrm{M}$] (m) at (5,1.5) {};
    \draw[thick, ->] (s) -- node[above] {$d_1$} ++ (m);

    \node[circle, fill = black, inner sep=0pt,minimum size=4pt] (r) at (3.161821,-0.05) {};

    % \draw[very thin, densely dashed] (r) -- (3.161821,1.72);
    % \draw[very thin, densely dashed] (3.161821,0.45) -- (3.161821,1.72);

    \draw[very thin] (s) -- node[below] {$d_2$} ++ (r);
    \draw[very thin, ->] (r) -- node[below] {$d_3$} ++ (m);

    \coordinate (origin) at (0,0);
    \coordinate (vert) at (3.161821,1.75);
    \coordinate (pivot) at (3.161821,0);
    \coordinate (sourcepoint) at (1,2);
    \coordinate (micpoint) at (5, 1.5);
    \coordinate (endpoint) at (7,0);

    % \pic [draw, "$\vartheta$", angle eccentricity=1.3] {angle = vert--pivot--sourcepoint};

    % \pic [draw, "$\vartheta$", angle eccentricity=1.3] {angle = micpoint--pivot--vert};

\end{tikzpicture}
    }
    \caption{Geometry of the simulated multi-path propagation in \emph{pyroadacoustics}: the microphone M receives the direct signal emitted by the source S, propagating via $d_1$, and the reflected sound, via $d_2$ and $d_3$~\cite{damiano_pyroadacoustics:_2022}.}
    \label{fig:pyroad_scene}
\end{figure}
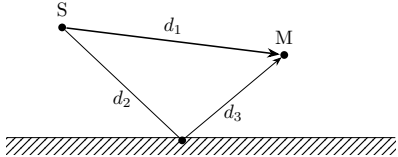

These motivations led to the design of~\emph{pyroadacoustics}, a simulator of sound propagation specifically targeting road scenarios~\cite{damiano_pyroadacoustics:_2022}, that has been developed during the first stage of the project and released as an open-source Python library~\cite{damiano_pyroadacoustics_2022}. The simulator has a flexible design that enables the user to adapt the acoustic scene parameters to different use cases. Its architecture, depicted in Fig.~\ref{fig:pyroad_architecture}, allows to simulate the sound produced by a single, omnidirectional sound source moving on an arbitrary trajectory with an arbitrary speed, and emitting a user-defined arbitrary audio signal. The sound is received by an array of an arbitrary number of omnidirectional, static microphones, placed in user-defined positions in the space, and includes both the direct component, and the reflection originating from the road surface (see Fig.~\ref{fig:pyroad_scene}). 
It should be noted that the support for user-defined trajectories ensures that the presence of both moving sources and microphones can be emulated by generating complex trajectories that account for variations in the relative speed between the source and the receivers (i.e. using spline or Bezier curves). To enhance the physical accuracy of the simulation, an accurate model of the reflection properties of the asphalt surface is included and implemented via finite impulse response (FIR) filtering, and can be adjusted by the user. Similarly, the effect of the air absorption is accounted for in the simulation via FIR filters. Finally, the use of variable-length delay lines to implement acoustic propagation ensures that the Doppler effect is correctly simulated~\cite{smith_physical_2010}. This tool is, to the best of the authors' knowledge, the only one publicly available providing the possibility to generate multichannel audio data in a road scenario, and can be exploited to obtain a dataset for both the detection and localization of sound sources, as well as to conduct studies on microphone array geometries for automotive applications.

% SDamiano: Is this paragraph really needed? It explains first steps towards detection, but does not add much novelty
In order to kick-start the design of the envisioned algorithms during the second stage of the project, \emph{pyroadacoustics} has then been used to generate a dataset for emergency vehicle detection. First, audio clips containing clean sounds of different types of sirens (namely, hi-low, wail and yelp~\cite{marchegiani_listening_2022}) and car horns have been collected from the online repository \emph{www.freesound.com}, together with $2.5$ hours of urban ambience and traffic noise.
All the audio events are recorded with a close microphone in order to avoid any interference from other sources active on the road. Exploiting these audio clips, a total of $15000$ single-channel data samples have been generated using \emph{pyroadacoustics}: each sample contains the sound emitted by a source of interest (i.e. either a siren or a car honking) moving on a random trajectory with an arbitrary speed, summed with a background noise extracted randomly from the collected noise clips.
%\sdamiano{each sample contains the sound emitted by a source of interest (i.e. either a siren or a car honking) moving on a random trajectory with a random speed, and a multi-layer background noise. To define it, multiple static and moving noise sources that emit traffic noise randomly extracted from the collected noise clips are randomly placed in the acoustic scene.} 
The sound and noise signals are combined with a random signal-to-noise ratio (SNR) in the range $[-30, 0]\,\si{\decibel}$. \sdamiano{While this dataset aims to fuel the exploration of sound event detection algorithms, the same simulator will be used for the generation of multi-microphone data to target localization. These tasks require to compare different architectures involving neural networks and traditional signal processing blocks (Sec.~\ref{sec:related_work}) and jointly optimize model complexity, robustness and accuracy.}
% \sdamiano{The generated dataset aims to fuel the development of algorithms for sound event detection in the automotive domain, requiring to explore different architectures (see Sec.~\ref{sec:related_work}) and jointly optimize model complexity, robustness and accuracy. Multi-microphone data will be then generated to explore microphone array processing techniques for noise reduction and localization, that could enhance the algorithmic performance and capabilities.}
\subsection{Hardware Flow Achievements}
\label{subsec:hardware_achievements}
\begin{figure}[t]
    \centering
    \includegraphics[width=0.8\linewidth]{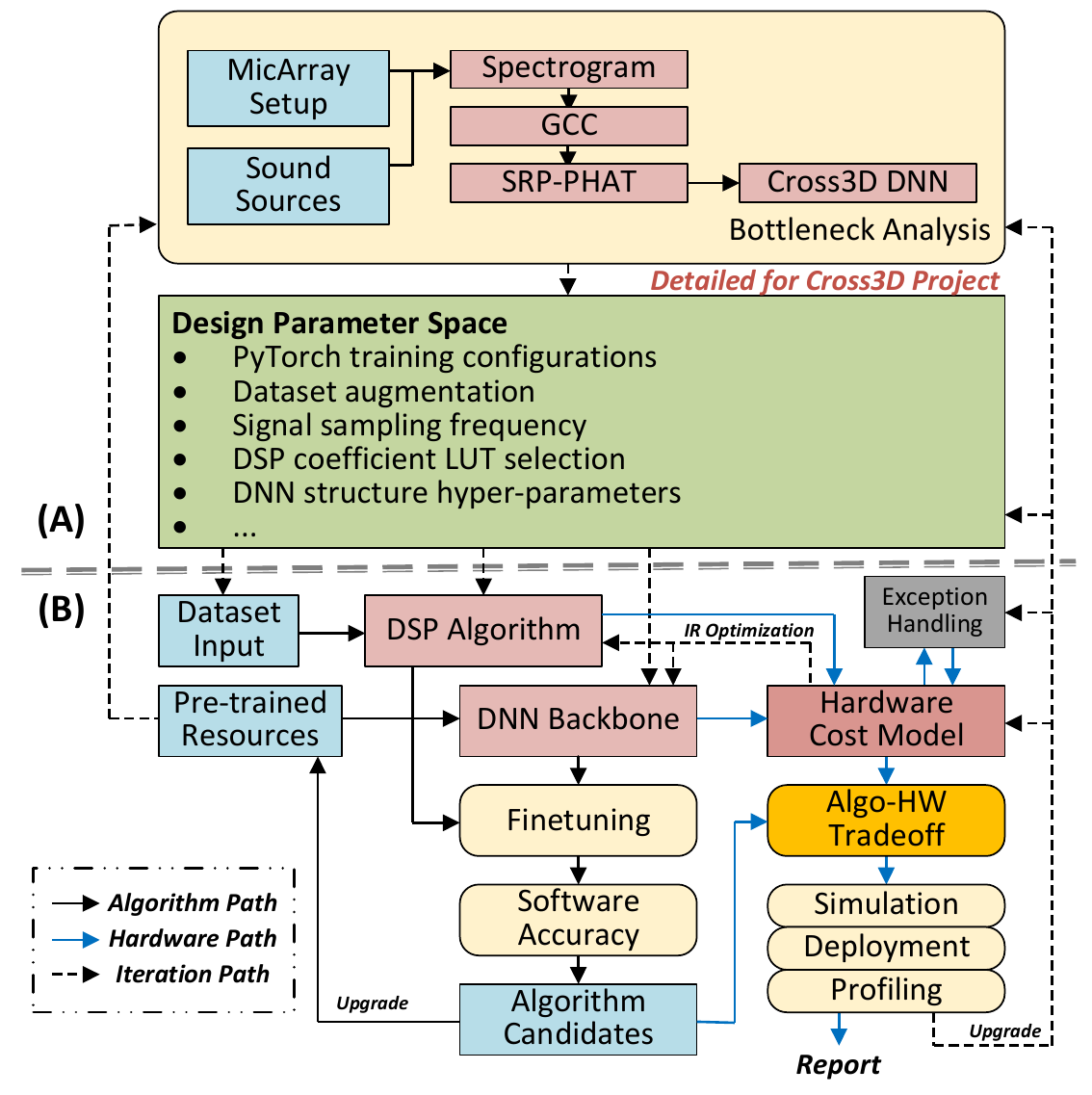}
    \caption{Overview of the hardware-algorithm co-design workflow prototype for I-SPOT hybrid algorithms, with (A) algorithm-specific analysis and (B) \review{}{general hardware-algorithm toolchain}.
    % \mv{[How can the algorithm agnostic part have blocks in which the accuracy or the SW trade-off is mentioned?]}. 
    The neural network part is powered by standard DNN frameworks. 
    The hardware profiling branch relies on IR porting from the original algorithm descriptions to unified lower operator expressions (currently with the TVM IR library \cite{chen2018tvm}).} 
    % \mv{[The legend talks about hardware Path, but the dotted lines are never used?]}
    \label{fig:hardware_cross3d_workflow}
\end{figure}

\jun{
% Preliminary studies on SSL algorithm optimization for hw accelerators
%As an initial attempt to facilitate the construction of 
To pipe-clean the hardware-algorithm development workflow, early-stage hardware-based bottleneck assessments and optimizations have been conducted %As a complement to the I-SPOT \review{}{dataset \emph{Pyroadacoustics}'s focus on sound events, the sound source localization (SSL) problem is tackled in this section with 
based on the Cross3D \cite{diaz-guerra_robust_2021} sound source localization (SSL) framework, a state-of-the-art baseline. Cross3D combines conventional signal pre-processing, with a deep learning back-end to achieve efficient and accurate SSL. It specifically
% dataset generator, a state-of-the-art DNN-based solution, Cross3D \cite{diaz-guerra_robust_2021}, has been selected which targets extremely randomized moving sound sources in the 3D space. The algorithm
\review{}{leverages pre-processing to extract the steered response power feature with phase transform filter (SRP-PHAT) and obtain robust SSL accuracy over unforeseen test data.} In the back-end, Cross3D replaces the typical hardware-unfriendly beam-forming computation for SRP-PHAT-based localization by a highly parallelizable CNN workload. In I-SPOT, we assessed the hardware benefits stemming from this optimization and further finetuned the Cross3D model towards more efficient edge deployment.}
To enable this, a hardware-oriented optimization workflow has been prototyped, as illustrated in Fig.~\ref{fig:hardware_cross3d_workflow}, covering the following key features:
\begin{enumerate*}[label=(\roman*)]
    \item The bottleneck analysis on the baseline algorithms across the design parameter space definition;
    \item \review{}{The algorithmic finetuning of design parameters based on resources including pre-trained model weights and empirical deep neural network (DNN) training parameters;}
    \item \review{}{The multi-level hardware cost model evaluation that jointly considers different design stages, such as roofline analysis \cite{williams2009roofline}, PyTorch profiler, and TVM runtime performance \cite{chen2018tvm});} % coherence optimization \mv{[what is a coherence optimization?]} to jointly 
    \item The hardware-algorithm trade-off judgment from algorithm output and hardware overhead;
    \item \review{}{The global configuration update to narrow the design parameter space and handle the co-design porting exceptions.}
    % \mv{Next iteration of what? To do what? What changes compared to the first iteration?}
    % \mv{[What do you mean with configuration update?]}
\end{enumerate*}

% Over multiple SRP resolution cases, the optimized model saves 10.32$\sim$73.71\% computational complexity and 59.77$\sim$94.66\% neural network weights against the Cross3D baseline, together with a better SRP implementation which halves the required complexity and on-chip memory overhead. As shown in Fig.~\ref{fig:hardware_cross3d_workflow}, the optimization is carried out in iteration with script-based workflow control, paving the road for future I-SPOT software-hardware co-design.

% \mv{[Can we add som very minimal results here? This makes it a bit more tangible, and will not kill you other publication...]}
This script-based workflow has proven its capability in squeezing Cross3D project for edge-device low-latency execution without giving in accuracy
\review{}{
(8.59 ms/frame end-to-end on RasPi-4B, 7.26x faster than the baseline). The hardware-based analysis and low-complexity SRP literature \cite{dietzen2021low} inspires a mathematically equivalent SRP-PHAT algorithm with $\sim$10x latency boost and $\sim$50\% coefficients reduce. The algorithm-hardware co-optimization helps to discover better training scripts and finetune the baseline model to edge-device versions which are $\sim$86\% smaller while $\sim$47\% faster.
}

\section{Open Challenges}
\label{sec:open_challenges}
Based on the achieved milestones in Section.~\ref{sec:achievements}, we are eager to tackle the remaining bottlenecks during the second stage of the project. On the algorithmic side, the design of sound event detection and localization algorithms targeting emergency sounds in a road scenario (i.e. emergency sirens and car horns) is currently ongoing and will be further investigated in the next few months. This task, supported by the dataset described in Sec.~\ref{sec:achievements}, will again combine both deep learning and traditional signal processing techniques, towards a hybrid approach to solve the SELD(t) problem. This hybrid approach aims at increased hardware efficiency and enhanced explainability of the results, a crucial aspect for a safety-critical use case as autonomous driving. 
Hybrid approaches, combining spectral and spatial features for signal analysis, have proven effective for the localization of sound events produced by moving sources, as discussed in Sec.~\ref{sec:project_goals}, but they are still rarely adopted in the automotive domain~\cite{furletov_auditory_2021}. 

\jun{
On the hardware side, two major challenges are to be tackled in the next project stage. 
First, the script-based co-design workflow requires extra functionality to enable higher automation: during algorithmic development, hybrid algorithms typically contain multi-level scientific computation application programming interfaces (APIs),
% \mv{in the form of Python library calls [or is this not what you mean? I do not understand what the Python reference comes doing here? Does it matter it is Python?]}
which differ in performance and hardware compatibility for porting and lowering. Hence, instead of handling these exceptions by manually selecting functions and rewriting the program towards better hardware mapping, an I-SPOT intermediate representation (IR) needs to be built. This will be pursued based on existing IR semantics, such as modifying the stateful data flow graph (SDFG) representation \cite{ben2019stateful} targeting heterogeneous CGRA fabric backends. 
\review{}{The toolchain prototype in Section.~\ref{subsec:hardware_achievements} also leaves space for further IR customization that lowers high-level algorithm descriptions to actually generate custom, yet flexible hardware.}
Second, the I-SPOT hardware architecture will be defined in more detail. Currently, the co-design optimization iteration ends with hardware evaluation on embedded CPU processors. Along with the progress and survey from the software aspect, the first version of CGRA processing elements and hardware control blocks will be drafted for basic operators in the target algorithm.
% Last, the designed algorithms will undergo an optimization stage prior to their deployment on the edge-devices currently being designed. The hardware assessment will drive the algorithmic exploration itself, via the exploitation of hardware-algorithm co-design techniques to include deployment requirements in an early stage of the design loop.
Moreover, with the flows currently in place, hardware design exploration and algorithm optimization will start to be more tightly interwoven. % in the second phase of I-SPOT.
}

Finally, on a system level, the design and assessment of the microphone array to be adopted for sensing purpose is yet to be tackled. This task also involves the assessment of the robustness of automotive SELD(t) methods to different microphone array geometries, justified by the need to potentially deploy these system on multiple classes of vehicles. The use of \emph{pyroadacoustics} to generate data with different sensor array architectures makes such an assessment feasible.

\section{Conclusion}
\label{sec:conclusion}
In this paper we introduced I-SPOT, an EU MSCA Project gathering two partner institutions, KU Leuven and Robert Bosch GmbH, that targets the enhancement of the acoustic perception of autonomous cars via audio signal processing. Within the project, the two aspects of algorithmic and hardware design co-exist in an interconnected development loop where hardware assessment requirements bring a foundational feedback into the algorithm design and tuning process, and vice versa.
During the first stage of the project a road acoustics simulator has been designed, setting the ground for both the development of emergency sound event detection and localization algorithms, and the design and assessment of the sensor array needed to provide the vehicle with sensing capabilities. 
The first version of a hardware-algorithm co-design flow has been established via evaluating and optimizing a state-of-the-art hybrid SSL solution, along with the initial data collection of hardware bottlenecks and overhead levels towards the I-SPOT Project requirements.
A hybrid approach has been chosen for the target audio signal processing algorithms, using % are based on 
a combination of traditional signal processing and deep learning techniques. This ensures an improved interpretability of the results compared to end-to-end deep learning methods, which is demanded by the safety-critical use case of autonomous driving. Remaining challenges are driving in our ongoing work and will be further addressed during the second stage of the project, together with (and with a continuous feedback from) the design of the edge-devices and accelerators on which the models will be deployed.

\section*{Acknowledgment}
This project has received funding from the European Union’s Horizon 2020
research and innovation programme under the Marie Skłodowska-Curie grant
agreement No. 956962 and from the European Research Council under the European Union's Horizon 2020 research and innovation program / ERC Consolidator Grant: SONORA (no. 773268). This paper reflects only the authors' views and the Union is not liable for any use that may be made of the contained information.

\bibliographystyle{ieeetr}
\bibliography{stefano, jun}

\end{document}